# Quantum Amplification on Molecular-Nuclear Transitions

V.B. Belyaev*, M.B. Miller**

* Joint Institute for Nuclear Research, 6 Joliot Curie, 141980 Dubna, Moscow region, Russia
e-mail: belyaev@theor.jinr.ru
**Institute in Physical-Technical Problems, 4 Kurchatova, 141980 Dubna, Moscow region, Russia
e-mail: mmiller@jinr.ru
tel: +7 49621 63596, fax: +7 49621 65863

Abstract

Molecular-nuclear transitions induced by external radiation are considered in this report, and new mechanism of quantum amplification based on this process is proposed. Two examples of the above effect are:

$\gamma + H_2O \rightarrow \begin{cases} {}^{18}Ne(1-;4.522) \\ {}^{17}F + p + 0{,}6MeV \end{cases}$, and $\gamma + {}^6LiD \rightarrow {}^8Be(2^+,0;\ 22{,}3) \rightarrow 2\alpha + 22{,}3MeV$.

## 1. Introduction: spontaneous molecular-nuclear transitions

Recently, the attention was paid to a possibility of molecular-nuclear transitions (MNT) in certain few-atomic molecules due to the presence of nuclear resonances in final nuclei (V. Belyaev and co-workers [1,2]). If the energies (and quantum numbers) of a molecular state and the nuclear resonance coincide, then the above molecule and nucleus can be considered as two degenerate states of the same quantum system. In this case, nuclear resonance wave functions have long tails, and hence noticeable overlap of molecular and nuclear wave functions is expected, and thus, a measurable admixture of nuclear state can takes place in corresponding molecules. The effect can show itself as spontaneous transitions between molecular and nuclear states: MNT. Following the work [2] the probability of spontaneous MNT in two-atomic molecule can be presented as $W(s^{-1})=\omega(E_1)\exp[0.6\pi\eta(E_2)]\cdot Q$ ,where $\omega(E_1)$ is a frequency, with which the nuclei in two-atomic molecule approach the barrier, $E_1$ is binding energy, and $Q$ is defined by the overlap integral between the wave functions of electronic configurations of the molecule and final atom, $1>Q>0.01$; $\eta(E_2)=Z_1Z_2e^2/hv$ is the parameter of Sommerfeld, where $v$ is the relative velocity of the outgoing particles defined by the energy of the transition between molecular and nuclear states of the system. This energy equals approximately to the nuclear resonance width.
Fig.1 shows these situations for $H_2O^*(1^-) \rightarrow {}^{18}Ne^*$ and ${}^6LiD \rightarrow {}^8Be^*$ systems.
Earlier, experiments on search for spontaneous MNT effect were carried out for these two molecular-nuclear combinations, and a lower limit of life-time for these cases of MNT was measured at a level $T_{1/2} \geq (3\div7)\cdot10^{19}$ y [3,4].

## 2. Induced molecular-nuclear transitions.

A necessary condition for spontaneous MNT is the identical equality of energy values of the resonant state of final atomic nucleus and that of initial molecular state. Obviously, this is a matter of chance, and it is impossible to control the situation in any way. In view of the experimental limitations, the number of objects for search is limited by a few molecular-nuclear combinations. In this connection, a stimulation of molecular-nuclear transitions by external radiation looks rather attractive [5].

Due to the uncertainties of experimental nuclear data it is not known whether the energies of the thresholds for a decay of nucleus via nucleon cluster channel are below or over the energies of the corresponding nuclear resonances. Let us consider a situation, when the molecular level is over the center of nuclear resonance, but still remains in a limit of the resonance width. Then, the molecular-nuclear complex represents a two level quantum system like it takes place in the quantum optics. However, there is a crucial difference between MNT and two-level quantum optical systems. Really, for the case of MNT no special procedure is necessary to create an active medium with inverse population. In the MNT case the initial molecular system in the ground (stable) state represents the active medium necessary for quantum amplification.

3. Analysis of induced MNT in terms of quantum coherent amplification

The considerations stated above let reasonable to consider the MNT under the external radiation (induced MNT) in terms of quantum coherent amplification.

For the quantum amplification, one can write a gain factor as $K=\exp(gL)$ [6], where $L$ is a linear size of the device, and $g$ is defined by a following expression:

$$g=N\cdot [\lambda^2 W_\gamma / 4(W_\gamma+\sum W_i)]. \qquad \text{(I)}$$

Here, $N=2.68\cdot 10^{19}$ cm$^{-3}$ is the Lochmidt constant, $\lambda$ – a wavelength of incident radiation, $W_\gamma$ is induced radiation probability (molecular-nuclear transition in our case), $\sum W_i$ is a sum of all possible channels of decay of the excited molecule different from the radiation. At the energy of several keV, which is typical for available cases of molecular-nuclear transitions, a main contribution to the sum $\sum W_i$ will be the value $W_B$, representing a probability of breakup of the molecule due to photo-dissociation.

The ratio $R=W_\gamma/4(W_\gamma+\sum W_i)$ can be estimated by comparing matrix elements defining probabilities $W_\gamma$ and $W_B$.

Radial part of matrix element defining probability $W_\gamma$ of $\gamma$ -transition with multipolarity $l$ can be written as

$$M_\gamma \sim \int {}^{J1}\Psi_{res}(r) r^l\, {}^{J2}\Psi_{mol}(r) r^2 dr\, ,\ l=|J1-J2|. \qquad \text{(II)}$$

Here, ${}^{J1}\Psi_{res}(r)$ and ${}^{J2}\Psi_{mol}(r)$ are wave functions of nuclear resonance and molecular states with angular moments $J1$ and $J2$, correspondingly.

When the energy of molecular state exceeds slightly the energy of nuclear resonance state, still remaining within the limits of the width of resonance, it is possible to expect a large overlap of wave functions in the integrand (II).

The above situation for $^6$LiD$\rightarrow$$^8$Be transition is illustrated in Fig.2. The molecular level of $^6$LiD supposed to be several keV over the $(2^+,0)$ nuclear resonance state of $^8$Be and located still in the limits of the width of this resonance, as it is shown in an inset at Fig.2.

At large distances between 6Li and D clusters, the wave function of (2+,0) resonance of $^8$Be behaves like the coulomb outgoing wave. Thus, the wave function of 8Be nucleus has a long tail. Due to this property of the wave functions, the matrix element (II) will be of the same order of magnitude as the matrix element of the breakup process MB, for which in the equation (II) one should use, instead of the resonance wave function Ψres(r), the wave function of a continuum spectrum. In that case one can expect that $M_B \sim M\gamma$, and, roughly speaking $R\approx 1/8$.

Now, we can estimate the gain-factor $g=N\cdot[\lambda^2 W_\gamma / 4(W_\gamma+\sum W_i)]$. Suppose the difference between two levels (molecular and nuclear) is $\Delta E$=hc/2$\pi\lambda$=1 keV. Here, h and c are Plank constant and the velocity of light, correspondingly. Then, for this case, we have:

$$g=2.68R\cdot 10^3 \text{cm}^{-1}. \qquad \text{(III)}$$

The dependence of $K=\exp(gL)$ on $R$ for $L$=1 cm is shown in Fig.2. It is seen that $K>1$ in wide range of $R$, what means that the amplification can take place.

## 4. Conclusion

Thus, irradiation of appropriate molecular system by electromagnetic radiation with energy in the range of kiloelectronvolt can induce the coherent molecular-nuclear transitions. Experimentally this phenomenon could be observed in coincidence measurements: one should look for the coincidences between pulses of keV-range primary radiation and radiation in MeV- range from the decay of highly excited final nuclei. For example, γ-quanta with energy $E_\gamma$=4.522 MeV will be emitted in the process $H_2O \rightarrow {}^{18}Ne$, and two α-particles with $E_\alpha$~11.2 MeV each are expected in ${}^6LiD \rightarrow {}^8Be$ case. Hence, this effect realizes a "molecular-nuclear amplification", in which the low energy electromagnetic radiation (X-rays or hard UV radiation) is "transformed" into the radiation with energy in MeV range: γ-quanta or α-particles. In principle, this effect can be considered as a new source of nuclear energy. In case of $H_2O \rightarrow {}^{18}Ne$ transition the energy yield is $\simeq$9.0 MeV, or $\simeq$0.5 MeV/nucleon, and for ${}^6LiD \rightarrow {}^8Be$ these values are 22.4 MeV and 2.8 MeV/nucleon, correspondingly, that is close in order of magnitude to the energy yield in nuclear fission of uranium or plutonium (~0.85 MeV/nucleon).

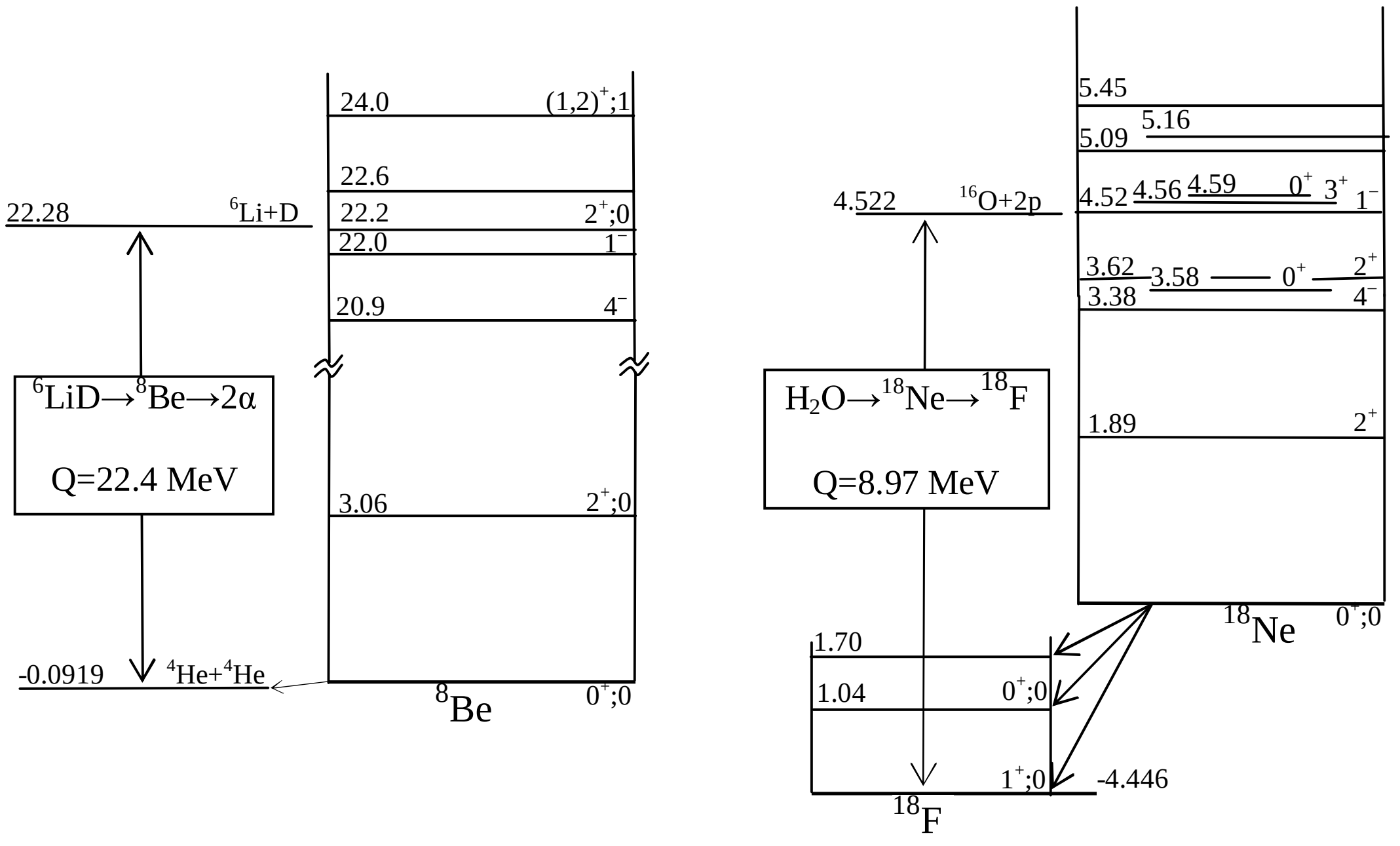

Fig.1. Fragments of nuclear level diagrams for $^{8}$Be and $^{18}$Ne.
Thresholds for few-body decay channels $^{6}$Li+d and $^{16}$O+2p in $^{8}$Be and $^{18}$Ne nuclei are indicated.

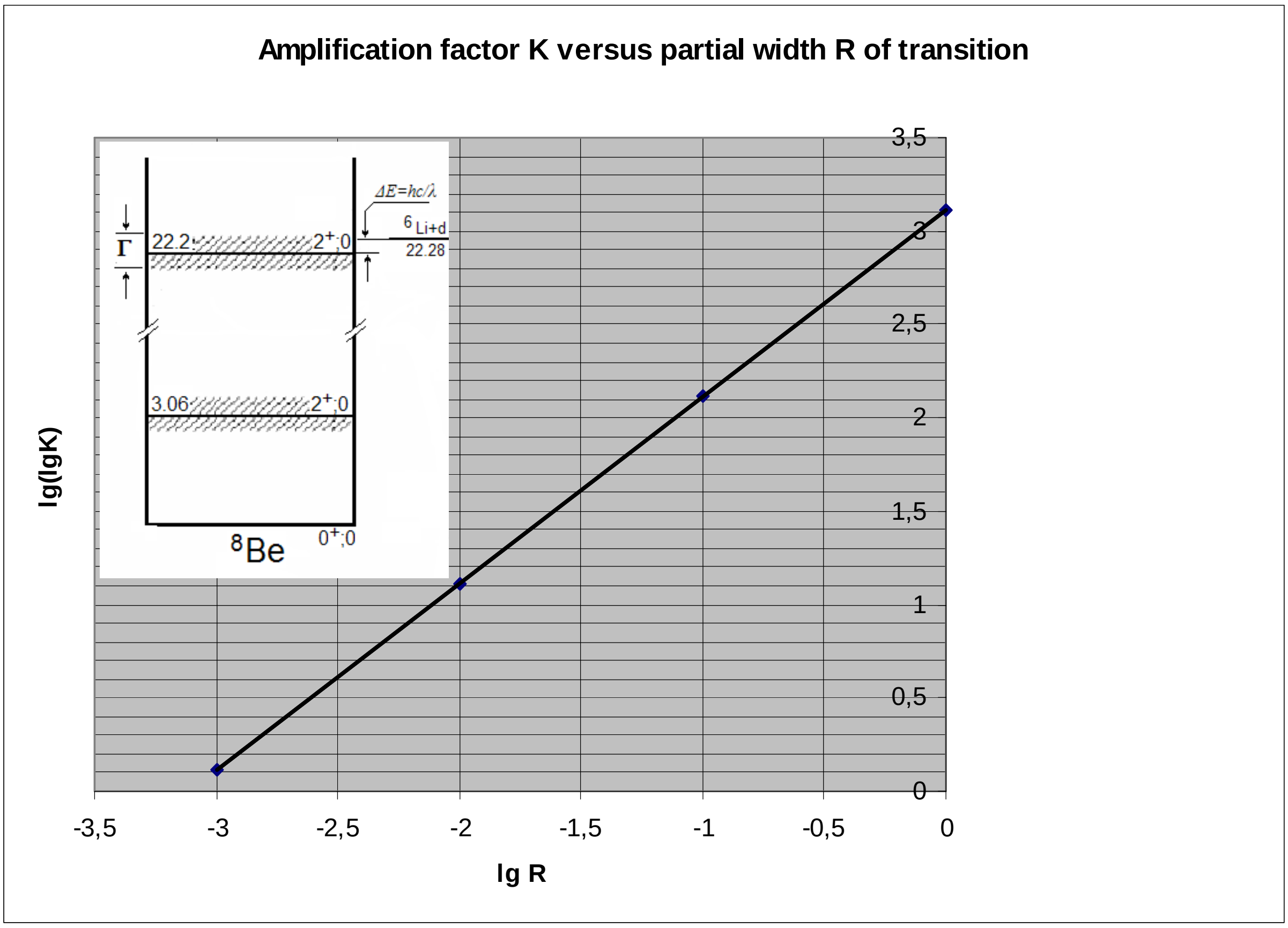

Fig.2. Gain factor versus the ratio *R*
In the insert, an idea of $^{6}$LiD→$^{8}$Be stimulated MNT is illustrated.